\documentclass{aa}
\usepackage[breaklinks=true,colorlinks,citecolor=blue]{hyperref}
\usepackage{units}
\usepackage{txfonts}
\usepackage{graphicx}
\usepackage{xcolor, tikz}
\bibpunct{(}{)}{;}{a}{}{,}

\newcommand{\modified}[1]{#1}  
\newcommand{\commented}[1]{}  

\newcommand*{\msun}{\mathit{M_\odot}}
\newcommand*{\zsun}{\mathit{Z_\odot}}
\newcommand*{\tms}{\mathit{t_\mathrm{MS}}}

\definecolor{lime}{HTML}{A6CE39}
\DeclareRobustCommand{\orcidicon}{%
    \begin{tikzpicture}
    \draw[lime, fill=lime] (0,0) 
    circle [radius=0.16] 
    node[white] {{\fontfamily{qag}\selectfont \tiny ID}};
    \draw[white, fill=white] (-0.0625,0.095) 
    circle [radius=0.007];
    \end{tikzpicture}
    \hspace{-2mm}
}

\defcitealias{2021ApJS..257...22S}{Paper I}
\defcitealias{2021ApJ...921..145S}{Paper II}

\graphicspath{{./}{figures/}}

\begin{document}

\title{Exploring the Stellar Rotation of Early-type Stars in the LAMOST Medium-resolution Survey. III. Evolution.}    
\titlerunning{Rotational evolution of early-type stars}
\author{Weijia Sun\href{https://orcid.org/0000-0002-3279-0233}{\orcidicon}\inst{1}\thanks{E-mail: wsun@aip.de} \and Cristina Chiappini\href{https://orcid.org/0000-0003-1269-7282}{\orcidicon}\inst{1,2}} 

\authorrunning{W. Sun \& C. Chiappini}

\institute{Leibniz-Institut für Astrophysik Potsdam (AIP), An der Sternwarte 16, 14482 Potsdam, Germany     \and{Laborat\'orio Interinstitucional de e-Astronomia - LIneA, Rua Gal. Jos\'e Cristino 77, Rio de Janeiro, RJ - 20921-400, Brazil}}
        
\date{Received ; accepted }
  
\abstract
   {Stellar rotation significantly shapes the evolution of massive stars, yet the interplay of mass and metallicity remains elusive, limiting our capacity to construct accurate stellar evolution models \modified{and to better estimate the impact of rotation in chemical evolution of galaxies}.}
   {Our goal is to investigate how mass and metallicity influence the rotational evolution of A-type stars on the main sequence (MS). We seek to identify deviations in rotational behaviors that could \modified{serve as new constraints to existing stellar models}.}
   {Using the LAMOST median-resolution survey Data Release 9, we derived stellar parameters for a population of 104,752 A-type stars. Our study focused on the evolution of surface rotational velocities and their dependence on mass and metallicity in 84,683 `normal' stars.}
   {Normalizing surface rotational velocities to Zero Age Main Sequence (ZAMS) values revealed a prevailing evolutionary profile from $1.7$ to $\unit[4.0]{\msun}$. This profile features an initial rapid acceleration until $t/\tms = 0.25\pm0.1$, potentially a second acceleration peak near $t/\tms = 0.55\pm0.1$ for stars heavier than $\unit[2.5]{\msun}$, followed by a steady decline and a `hook' feature at the end. Surpassing theoretical expectations, the initial acceleration likely stems from a concentrated distribution of angular momentum at ZAMS, resulting in a prolonged increase in speed. A transition phase for stars with $2.0<M/\msun<2.3$ emerged, a region where evolutionary tracks \modified{remain uncertain}. Stellar expansion primarily drives the spin down in the latter half of the MS, accompanied by significant influence by inverse meridional circulation. The inverse circulation becomes more efficient at lower metallicity, explaining the correlation of the slope of this deceleration phase with metallicity from $\unit[-0.3]{dex}$ up to $\unit[0.1]{dex}$. Starting with lower velocities at ZAMS, the metal-poor ($\unit[-0.3]{dex} < \mathrm{[M/H]} < \unit[-0.1]{dex}$) subsample suggests a mechanism dependent on metallicity for removing angular momentum during star formation. The proportion of fast rotators decreases with an increase in metallicity, up to $\log(Z/\zsun)\sim -0.2$, a trend consistent with observations of OB-type stars found in the Small and Large Magellanic Clouds.
	}
   {}

\maketitle
  
\section{Introduction}
\label{sec:intro}

Stellar rotation is a crucial factor in the evolution of massive stars, significantly affecting their structure through geometrical distortion \citep{2012A&ARv..20...51V}, extra-mixing \citep{2001MNRAS.327..353H}, and enhanced mass loss rates \citep{2008A&ARv..16..209P}. Hydrodynamical instabilities induced by meridional circulation \citep{1992A&A...265..115Z} lead to turbulent motions, which facilitate the mixing of chemical compositions \citep{2010A&A...522A..10C} and the redistribution of internal angular momentum \citep{2019ARA&A..57...35A}.

While extensive efforts have been devoted to modeling the effects of rotation \citep[e.g.,][]{2000ARA&A..38..143M, 2000ApJ...544.1016H, 2008A&A...478..467E, 2018ApJS..237...13L}, the internal transport of angular momentum remains a significant challenge \citep{2023Galax..11...54Z}. This issue is central to our understanding of stellar evolution, as the distribution of angular momentum directly influences the evolutionary outcomes. The literature reflects a diversity of approaches to diffusive and advective transport mechanisms \citep{2000ARA&A..38..143M, 2000ApJ...544.1016H, 2012MNRAS.419..748P}, compounded by assumptions on the initial rotation law at the Zero Age Main Sequence (ZAMS), which carries important consequences for the further evolution of stars on the main sequence (MS). Despite considerable progress over the last decade, the complexity of this research is further amplified by the consideration of magnetic fields \citep{2012ApJ...748...97R, 2018MNRAS.477.2298Q, 2021A&A...646A..19T}, early accretion history \citep{2017A&A...602A..17H}, {and binarity \citep{2012Sci...337..444S, 2013ApJ...764..166D, 2021A&A...653A.144H}.}

Addressing these pivotal challenges has led to the development of innovative methodologies. Since the proposal by \citet{1980Ap&SS..73..159A} to estimate the inner rotational angular velocity based on the rotational frequency splitting of non-radial pulsations, asteroseismic data has significantly propelled the observation of internal rotation in stars beyond the Sun \citep[{e.g.,}][]{2012A&A...539A..63R, 2015A&A...583A..62B, 2021ApJS..255...17S, 2023A&A...673L..11B}. This method has provided a unique opportunity to probe the interior, albeit its application is confined to specific types of hot stars such as slowly pulsating B stars \citep{2022ApJ...940...49P} and $\gamma$ Doradus stars \citep{2019MNRAS.487..782L}. In parallel, the analysis of surface rotational velocities derived from spectroscopic data has shed light on the complex processes of global angular momentum loss and redistribution within stars, illuminating the respective efficiencies of these competing processes \citep{2007A&A...463..671R}. \citet{1970stro.coll..207D} suggested that Maxwellian distributions of rotational velocities could indicate a state of rigid rotation established at the ZAMS. Contrary to this expectation, studies, including \citet{1995ApJS...99..135A}, discovered a bimodal distribution of rotational velocities among A-type MS stars, arguing that slower rotations might predispose stars to develop chemical peculiar (CP) stars. Further examination by \citet{2007A&A...463..671R} confirmed the presence of genuine bimodal distributions in the equatorial rotational velocities by excluding CP stars in their sample, with variations reflecting differences among spectral types. Expanding on this framework, \citet{2012A&A...537A.120Z} established rotational distribution models and evolutionary maps for stars with masses around $\unit[2-3]{\msun}$, based on data from two thousand early-type field stars. This analysis revealed a pronounced acceleration in rotational velocity within the first third of the MS evolutionary phase, a phenomenon that existing rotation models did not predict.

Validating theoretical models requires extensive samples across a broad spectrum of stellar atmospheric parameters, achievable only through large spectroscopic surveys like the Large Sky Area Multi-Object Fiber Spectroscopic Telescope \citep[LAMOST,][]{2012RAA....12.1197C, 2020arXiv200507210L} and the upcoming 4-metre Multi-Object Spectroscopic Telescope \citep[4MOST,][]{2019Msngr.175....3D}. While the \textit{Gaia} \citep{2016A&A...595A...1G, 2023A&A...674A...1G} mission has conducted the largest survey for rotational line-broadening measurements to date in its third data release (DR), it suffers from Radial Velocity Spectrometer's narrow wavelength range, affecting sensitivity to line broadening in hot stars \citep{2023A&A...674A...8F}. A modified version of \textit{The Payne} was used by \citet{2022A&A...662A..66X} to derive stellar atmospheric parameters, including the projected rotational velocity $v\sin i$, for 330,000 OBA-type stars in LAMOST low-resolution ($R\sim 1800$) survey (LRS) DR 6. Furthermore, \citet[hereafter Paper I]{2021ApJS..257...22S} leveraged LAMOST's median-resolution ($R\sim 7500$) survey (MRS) DR 7 to catalog over 40,000 late-B and A-type MS stars, uncovering a bimodal rotation distribution that varies with stellar mass and metallicity. Specifically, metal-poor stars {($\unit[-1.3]{dex}<\mathrm{[M/H]} < \unit[-0.2]{dex}$, with an average \modified{value} around $\unit[-0.36]{dex}$)} predominantly show slow rotation, whereas metal-rich stars {($\unit[-0.2]{dex}<\mathrm{[M/H]}<\unit[0.69]{dex} $, with an average \modified{value} around $\unit[0.32]{dex}$)} exhibit both slow and rapid rotational branches \citep[hereafter Paper II]{2021ApJ...921..145S}.

This study expands on the investigations by \citetalias{2021ApJS..257...22S} and \citetalias{2021ApJ...921..145S}, utilizing the significantly larger dataset from LAMOST MRS DR 9. By doubling the sample size, we {study} deeper into the evolution of stellar rotation on the MS, particularly how it varies with mass and metallicity. Our analysis includes a critical comparison with theoretical rotation models to dissect the processes of angular momentum's initial distribution and its subsequent redistribution. Moreover, we investigate the fraction of rapid rotation among different metallicity levels and draw comparisons with analogous phenomena in more massive stars within dwarf galaxies to explore how metallicity affects rotational evolution.

This article unfolds as follows. Section~\ref{sec:data} outlines our methodology for data reduction and contamination cleaning. In Section~\ref{sec:result}, we examine the evolution of stellar rotation, analyzing its dependence on mass and metallicity, and compare these findings with observations of rapid rotators in dwarf galaxies of the Magellanic Clouds. Section~\ref{sec:discussion} {looks} into the mechanisms influencing rotation evolution, exploring observed phenomena relative to existing stellar rotation models, and discusses metallicity's role and the study's limitations. Section~\ref{sec:summary} concludes with a summary of our key findings and implications.

\section{Data}
\label{sec:data}
We followed the candidate selection and data reduction procedures for MRS in LAMOST DR9 as established in \citetalias{2021ApJS..257...22S}. Specifically, we selected coadded spectra from the LAMOST MRS General Catalog, excluding objects with effective temperatures below {\unit[5000]{K}} (as determined by the LAMOST Stellar Parameter pipeline, LASP) or those with a median pixel signal-to-noise ratio (S/N) less than 15. {The average temperature uncertainty reported by LASP for the remaining sample is around \unit[40]{K}.} We then cross-matched our candidates with the Two Micron All Sky Survey \citep[2MASS;][]{2006AJ....131.1163S} for extinction correction and with Gaia DR3 \citep{2016A&A...595A...1G, 2023A&A...674A...1G} for photometric determination, adhering to \citet{2021A&A...649A...3R}'s guidelines on Gaia astrometric and photometric parameters. {We adopted the photogeometric distance estimates of \citet{2021AJ....161..147B}.} \commented{The reason why I did not use the StarHorse is that most of these stars are not included in the LAMOST Stellar Parameter Catalog (no temperature), so they are also missing in the StarHorse.}

To further refine our selection, we used line indices of H$\alpha$ (\unit[6548.0-6578.0]{\AA}) and Mg\,{\sc i} \textit{b} (\unit[5160.12--5192.62]{\AA}), which are robust against extinction correction and flux calibration issues \citep{2015RAA....15.1137L}. Only MRS samples with line-indices-based temperatures above \unit[7000]{K} were considered. Spectra normalization and transformation to the rest frame were performed using the \texttt{laspec} package \citep{2021ApJS..256...14Z}. The Stellar LAbel Machine \citep[SLAM,][]{2020ApJS..246....9Z}, trained on ATLAS12 atmospheric models \citep{2005MSAIS...8...25C}, were used to infer the stellar labels of $T_\mathrm{eff}$, $\log g$, $\mathrm{[M/H]}$ and $v\sin i$. Our model grid spanned a range of $T_\mathrm{eff}$ from \unit[6000]{K} to \unit[15,000]{K}, $\log g$ from \unit[3.5]{dex} to \unit[4.5]{dex}, $\lbrack\mathrm{Fe}/\mathrm{H}\rbrack$ from $\unit[-1.0]{dex}$ to $\unit[1.0]{dex}$, and $v\sin i$ from $\unit[10]{km,s^{-1}}$ to $\unit[500]{km,s^{-1}}$ \citepalias[for more details see][]{2021ApJS..257...22S}. We confirmed there is no systematic bias against the $T_\mathrm{eff}$.

\begin{figure}[ht!]
\centering
\includegraphics{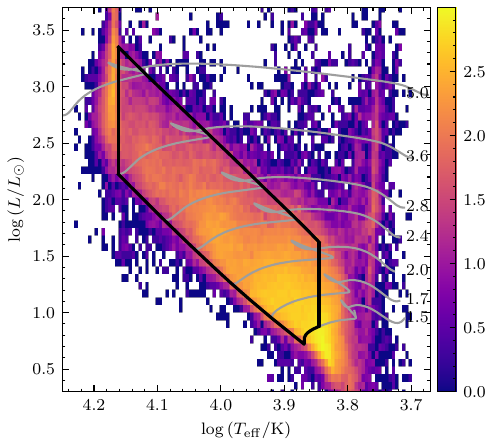}
\caption{Hertzsprung--Russell diagram for the full sample. Evolutionary tracks for stars of masses 1.5, 1.7, 2.0, 2.4, 2.8, 3.6, and $\unit[5]{\msun}$ are illustrated as gray curves. Black lines indicate selection boundaries for stars with $M/\msun \geqslant 1.5$ of solar metallicity, incorporating vertical temperature limits at \unit[14500]{K} and \unit[7000]{K}. Color gradients denote stellar density in each bin on a logarithmic scale. {The spikes on the hot ($T_\mathrm{eff}\sim \unit[15000]{K}$) and cool ($T_\mathrm{eff}\sim \unit[6000]{K}$) ends correspond to the temperature limits of the training models.}\label{fig:hr}}
\end{figure}

\begin{figure}[ht!]
\centering
\includegraphics{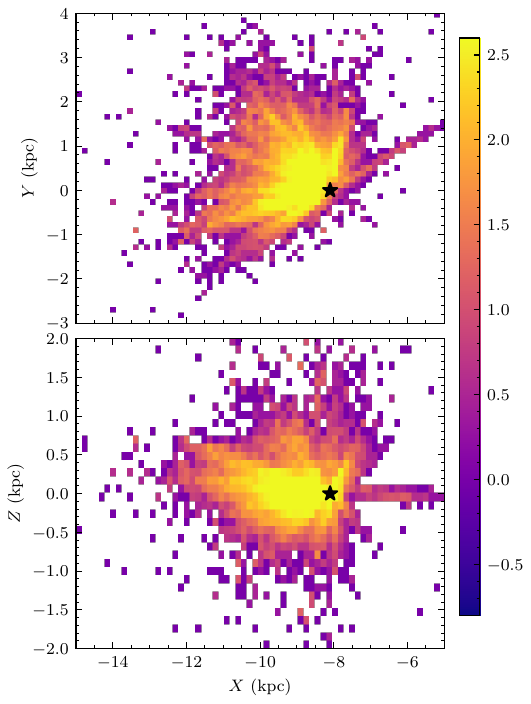}
\caption{Projection of stellar locations on the galactic $X - Y$ plane (top) and $X - Z$ plane (bottom). The position of the Sun is marked by a black star symbol for reference. Color gradients indicate the density of stars in each spatial bin, quantified on a logarithmic scale. \label{fig:gal}}
\end{figure}

In Figure~\ref{fig:hr}, we present the Hertzsprung--Russell (HR) diagram for our complete dataset. To mitigate artifacts from our model's temperature limitations, we implemented an upper temperature threshold of \unit[14500]{K} (indicated by black vertical lines), thereby excluding an anomalous concentration of stars {beyond or close to our training grids of temperature}. Similarly, a lower temperature limit of \unit[7000]{K} was applied to filter out low-mass stars from our analysis. Additionally, stars that have evolved beyond the MS were excluded from this selection process. To enhance the reliability of our projected rotational velocity ($v\sin i$) measurements, we further removed any star with a relative uncertainty $\sigma_{v\sin i}/v\sin i > 0.05$. \modified{With a median S/N around 60, the cross validated scatter for $T_\mathrm{eff}$, $\log g$, [M/H], and $v\sin i$ are $\sim\unit[75]{K}$, $\unit[0.06]{dex}$, $\unit[0.05]{dex}$, and $\sim\unit[3.5]{km\,s^{-1}}$, respectively \citepalias{2021ApJS..257...22S}.} Following these criteria, our refined catalog comprises 104,752 stars, whose spatial distribution is shown in Figure~\ref{fig:gal}. \modified{This distribution in the disk largely follows the subset described in \citetalias{2021ApJS..257...22S}, as shown in Figure~1 of \citet{2023A&A...674A.129W}. Moreover, \citet{2023A&A...674A.129W} looked into the radial metallicity gradients and azimuthal metallicity distributions on the Galactocentric $X$–$Y$ plane using mono-temperature stellar populations of \citetalias{2021ApJS..257...22S}, and found negative metallicity gradient ranges decreasing as the effective temperature decreases.}

\subsection{Contamination}
\label{sec:contamination}
To ensure our sample reflects intrinsic stellar rotation unaffected by external factors, we identified four primary contamination types: CP stars, binaries, cluster members, and periodic variables. CP stars are early-type MS stars with unusual chemical compositions, often linked to slower rotation speeds under $\unit[120]{km,s^{-1}}$ \citep{1974ARA&A..12..257P, 1995ApJS...99..135A}. Close binaries can experience altered rotation due to tidal forces, mass transfer, or mergers, affecting the rotational characteristics of both components \citep{2013ApJ...764..166D}. Cluster members may exhibit rotational velocity distributions that diverge from field stars, reflecting their unique formation and evolutionary histories \citep{2006ApJ...648..580H}.

In alignment with \citetalias{2021ApJ...921..145S}, we assessed the relative variation in radial velocity from single-exposure spectra via cross-correlation, excluding stars with a variation exceeding $\unit[4]{\sigma}$ and an absolute change greater than $\unit[16]{km,s^{-1}}$. This criterion identified 15,527 stars as spectroscopic binaries, which represent approximately 15\% of our sample. {A tentative trend is found between binary fraction and metallicity, wherein the fraction decreases from 18\% to 13\% as the mean metallicity increases from approximately $\unit[-0.5]{dex}$ to $\unit[0.05]{dex}$, similar to the close binary fraction of solar-type stars \citep{2010ApJS..190....1R, 2019ApJ...875...61M}.} For additional contamination types, we cross-matched our data with pre-compiled catalogs from \citet{2019ApJS..242...13Q}, \citet{2020A&A...640A..40H}, \citet{2021A&A...645A..34P}, and \citet{2023ApJS..266...14T} to pinpoint CP stars, and with \citet{2020ApJS..249...18C} for periodic variables. We also utilized an updated catalog \citep{2023A&A...673A.114H} based on Gaia DR3 to identify cluster members. This process resulted in the identification of 2,765 CP stars, 886 periodic variables, and 1,952 cluster members. \modified{Among the periodic variables with type classification, over 85\% are labeled as EW-type, EA-type eclipsing binaries, and RS Canum Venaticorum–type systems. Non-binary systems like $\delta$ Scuti account for less than 10\%. This indicates that the majority of these variable contaminants are likely in multi-star systems, which need to be excluded from further analysis. While $\delta$ Scuti might be considered `normal' in this sense, these detected variables may exhibit variability in luminosity and spectral features, showing an average amplitude variation of \unit[0.15]{mag} in the Zwicky Transient Facility $g$ band. This level of variability indicates that single-epoch observations without phase information are insufficient to accurately characterize these stars, thereby affecting the precision of parameter measurements. Consequently, we determined that their exclusion is warranted.} After these exclusions, we categorized the remaining 84,683 stars as our `normal' sample, comprising single, non-CP, non-variable stars not affiliated with clusters. {The numbers of different \modified{contaminants} are tabulated in Table~\ref{tbl:catalog}.} \modified{CP stars of different subgroups are also tabulated: metallic line (Am), magnetically peculiar (mAp), stars with enhanced Hg {\sc ii} and Mn {\sc ii} (HgMn), and He-weak stars.}

While we attempted to identify and exclude contaminations, it is possible that some undetected contamination remains. The detection of cluster members is considered complete, yet other methods face challenges. Specifically, our binary-detection technique lacks sensitivity to wide binaries, with effectiveness further constrained by the time cadence and observation frequency. For CP stars, detection rates reported by \citet{2019ApJS..242...13Q} are significantly lower than those found in previous studies \citep{1974ARA&A..12..257P, 1981ApJS...45..437A, 2012A&A...537A.120Z}, indicating potential underestimation. Additionally, the selection in \citet{2019ApJS..242...13Q}, based on LAMOST DR5's Low-Resolution Survey, covered approximately 70\% of LRS targets and about 25\% of the MRS targets in DR9, suggesting a coverage gap that could influence detection completeness.

\begin{table}
      \caption[]{Number of sources in catalogues.}
         \label{tbl:catalog}
     $$
         \begin{array}{p{0.5\linewidth}r}
            \hline
            \noalign{\smallskip}
            Catalogue      &  N \\
            \noalign{\smallskip}
            \hline \hline
Total stars                                            & 104,752 \\ \hline
Contaminants                                           &         20,069 \\ \hline
\hspace*{0.5cm} Binary stars                                           & 15,527  \\
\hspace*{0.5cm}Cluster members                                        & 1,952   \\
\hspace*{0.5cm}Variable stars                                         & 886     \\
\hspace*{0.5cm}CP stars                                               & 2,765   \\ 
\hspace*{1.0cm}\modified{Am}                                               & 2,690   \\ 
\hspace*{1.0cm}\modified{mAp}                                               & 233  \\ 
\hspace*{1.0cm}\modified{HgMn}                                               & 22   \\ 
\hspace*{1.0cm}\modified{He-weak}                                               & 1   \\ \hline
\hline
Normal stars                                 & 84,683  \\ \hline
\hspace*{0.5cm}$-1.0 < \mathrm{[M/H]} \leqslant -0.3$ & 8,972   \\
\hspace*{0.5cm}$-0.3 < \mathrm{[M/H]} \leqslant -0.1$   & 19,463  \\
\hspace*{0.5cm}$-0.1 < \mathrm{[M/H]} \leqslant 0.0$   & 16,216  \\
\hspace*{0.5cm}\hspace*{0.5cm}$0 < \mathrm{[M/H]} \leqslant 0.1$   & 15,589 \\
            \noalign{\smallskip}
            \hline
         \end{array}
     $$
\end{table}

\subsection{Stellar mass and age}
\label{sec:mass}
We estimated the mass and age of stars in our sample using the Stellar Parameters INferred Systematically (SPInS) tool \citep{2020A&A...642A..88L}, a Bayesian framework designed to derive the probability distribution functions of stellar parameters. This analysis incorporated stellar evolution models from the Bag of Stellar Tracks and Isochrones \citep[BASTI,][]{2004ApJ...612..168P, 2006ApJ...642..797P} and adopted the Kroupa initial mass function \citep{2001MNRAS.322..231K, 2013pss5.book..115K} as a prior. Inputs for SPInS included: (1) effective temperature as determined by SLAM; (2) bolometric luminosity, calculated from Gaia parallaxes and 2MASS $K$-band magnitudes, corrected for extinction and bolometric corrections; and (3) metallicity as derived by SLAM. The inference of stellar mass, age, and radius was performed using a Markov chain Monte Carlo method.

\begin{figure}[ht!]
\centering
\includegraphics{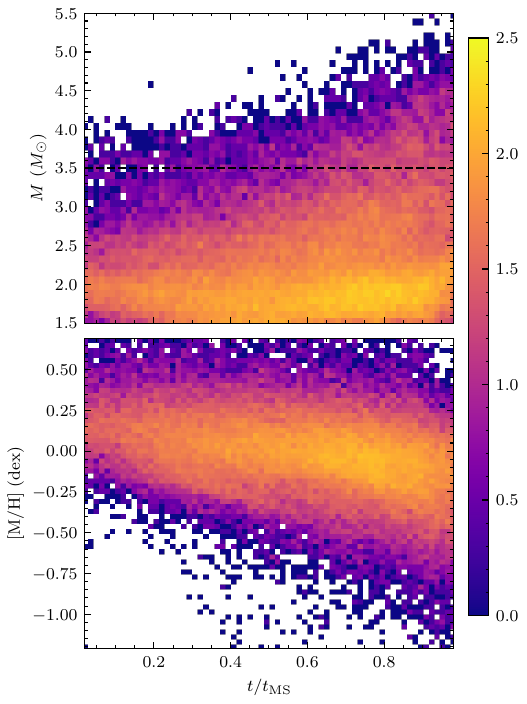}
\caption{Age ($t/\tms$) -- mass distribution (top) and age -- metallicity distribution (bottom) of the selected normal stars. {A horizontal dashed line is overplotted at $M=\unit[3.5]{\msun}$, beyond which young stars are excluded due to the temperature cut at $T_\mathrm{eff}=\unit[14500]{K}$.} \label{fig:prop}}
\end{figure}

We translated stellar ages into their respective evolutionary phases on the MS, represented as the ratio $t/\tms$, where $t$ denotes the age of a star since the ZAMS, and $\tms$ signifies the duration from ZAMS to the Terminal Age Main Sequence (TAMS). Figure~\ref{fig:prop} illustrates the distribution across the $t/\tms$--$M$ (mass) plane (top panel) and $t/\tms$--$\mathrm{[M/H]}$ (metallicity) plane (bottom panel) for stars categorized as `normal'. Our sample's distribution within the mass--age diagram aligns closely with the findings of \citetalias{2021ApJS..257...22S}, illustrating consistent profiles across studies. The noticeable absence of younger stars with masses exceeding $\unit[3.5]{\msun}$ can be attributed to the implemented temperature cut at $T_\mathrm{eff}=\unit[14500]{K}$, as marked by the vertical line in Figure~\ref{fig:hr}, which systematically excludes these higher mass, potentially hotter stars from our analysis. Furthermore, an analysis of metallicity trends reveals a gradual shift towards an older stellar population as metallicity decreases, suggesting that metal-poor stars in our sample tend to be further along in their evolutionary phases. \modified{The median uncertainty of the mass and age are $\unit[0.03]{\msun}$ and $\unit[0.03]{\tms}$.}

\section{Result}
\label{sec:result}
In this section, we undertake a detailed investigation into the dynamics of stellar rotation, analyzing how it evolves and its dependence on stellar mass and metallicity. Recognizing the challenge posed by the projection effect in measuring spectroscopic rotational velocity, we employed a binning method. Within each bin, we calculated the average projected rotational velocity, $\langle v\sin i\rangle$, subsequently converting these averages to true rotational velocities, $\langle v\rangle$. This conversion relies on the assumption of stars having randomly oriented rotational axes, a premise supported by the statistical framework outlined by \citet{1950ApJ...111..142C}. {Note that average values of $v\sin i$ might not well reflect a population's rotational evolution if the underlying distribution is complicated (see discussion in Section~\ref{sec:caveat}).}

We initiate our analysis by examining the evolution of rotational velocity as a function of stellar mass, focusing on the ratio $\langle v/v_\mathrm{ZAMS}\rangle$ across different mass categories. This first step allows us to isolate the effect of mass on rotational dynamics, setting a foundation for further exploration into how metallicity influences these patterns. {We then compared this ratio against the predictions from models, ensuring that both values begin at unity at the onset of the MS and evolve on the same scale. While the absolute rotational velocity is largely determined by the initial angular momentum, the adopted initial rotation rates in the models have a minor impact on this ratio. Consequently, the corresponding evolutionary profile exhibits a similar behavior that remains relatively stable, as we will see in Section~\ref{sec:evolution}.} By categorizing our sample into distinct metallicity bins, we then explore how metallicity affects stellar rotation, providing insight into the complex interplay between mass, age, and metallicity. This gradual refinement from mass to metallicity is followed by a broader comparison with stellar populations in different galaxies. We analyze the prevalence of rapid rotators within these populations as a function of metallicity, offering a holistic view of the varied forces influencing stellar rotation across diverse galactic environments.

For the majority of our analysis, we focus on stars with metallicity $\mathrm{[M/H]}$ less than \unit[0.1]{dex}, except for discussions in Sections~\ref{sec:rapid} where we extend our considerations to include a broader range of metallicity. This decision is informed by the complexity of modeling stars with super-solar metallicity, where the increased abundance of heavy elements leads to enhanced opacity, affecting stellar winds, mass loss, and convective overshooting \citep[e.g.,][]{2006astro.ph.11261M}, as highlighted by observations in metal-rich environments like the Galactic center or metal-rich globular clusters \citep[e.g.,][]{2022MNRAS.511.2814Y}. We also introduce a lower metallicity limit of $\unit[-0.3]{dex}$ for specific analyses in Section~\ref{sec:evolution} and \ref{sec:evolution_metal}, ensuring a focused examination of stellar rotation within a well-defined parameter space. {The numbers of stars within the metallicity bins used in this work are tabulated in Table~\ref{tbl:catalog}.}

\subsection{Evolution of rotation as a function of mass}
\label{sec:evolution}


\begin{figure*}[ht!]
\centering
\includegraphics{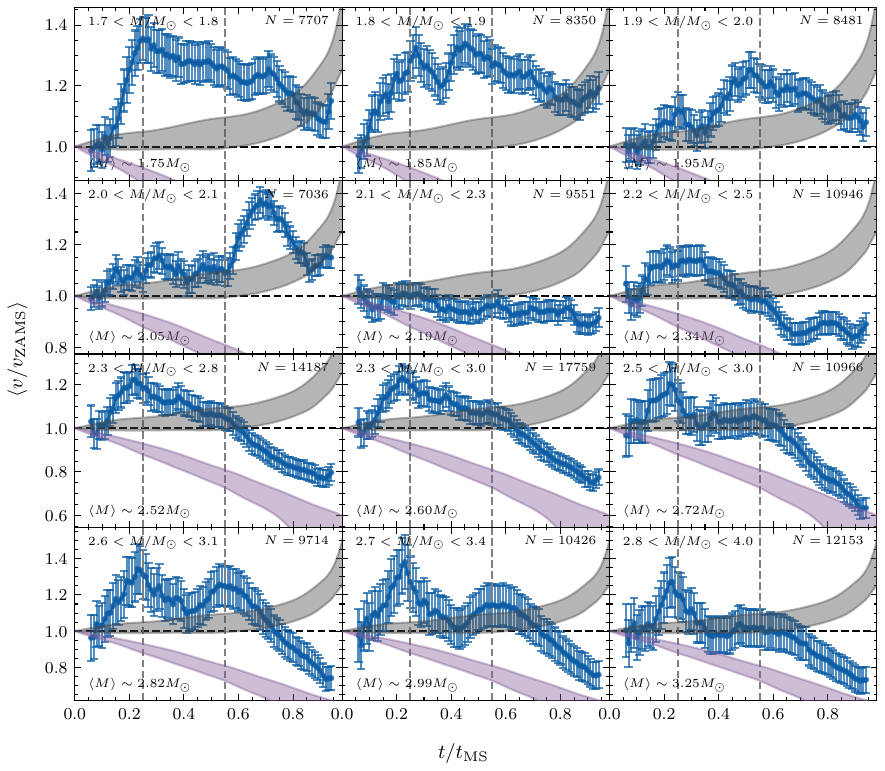}
\caption{Evolution of the equatorial rotational velocity ratio $\langle v/v_\mathrm{ZAMS}\rangle$ throughout the MS lifespan, across various mass ranges. $v_\mathrm{ZAMS}$ refers to the initial equatorial velocity at the ZAMS. The blue curves represent the average rotation rates, obtained using a sliding average over age, with a window size of $\unit[0.1]{\tms}$, and error bars representing uncertainties estimated via bootstrapping. The average mass for each bin is indicated in the bottom-left corner \modified{and the number of targets is shown in the top-right corner}. Shaded regions illustrate the theoretical evolution of the $v/v_\mathrm{ZAMS}$ ratios for rigid rotators ({gray}) and for simple differential rotators ({purple}). Vertical dashed gray lines at $t/\tms=0.25$ and 0.55, respectively, highlight significant phases within the MS lifespan.
 \label{fig:vevo_zams_mass}}
\end{figure*}

In Figure~\ref{fig:vevo_zams_mass}, we explore the equatorial velocity ratio $v/v_\mathrm{ZAMS}$ across the MS lifespan for stars within mass ranges from $1.75$ to $\unit[3.25]{\msun}$. $\langle v/v_\mathrm{ZAMS}\rangle$ is calculated utilizing a sliding average across $t/\tms$, with a bin width of 0.1. We confirm that the choice of sliding window size and mass binning minorly influences the observed trends. We overplot the observed velocities against theoretical expectations for two distinct rotational models: stars evolving as rigid rotators {(gray)}, characterized by $\langle v/v_\mathrm{ZAMS}\rangle>1$, and those as differential rotators {(purple)}, which do not redistribute angular momentum and have $\langle v/v_\mathrm{ZAMS}\rangle<1$. The shaded areas depict theoretical evolutionary paths for mass regimes ranging from $1.5$ to $\unit[3.0]{\msun}$, as {discussed} in \citet[see their Figure~10]{2012A&A...537A.120Z}. According to these models, rigid rotators exhibit a gradual increase in rotational velocity with age, remaining relatively constant through the initial two-thirds of the MS phase before significantly increasing in the final third. This acceleration is attributed to a mass-compensation effect, as described by \citet{1970A&A.....8...76S}, where rotation enhances central density and facilitates the transfer of angular momentum from the core to the surface. Conversely, for differential rotators, changes in surface velocity result from an expanding radius as stars progress from ZAMS to TAMS.

The observed evolution of rotational velocity presents notable discrepancies from theoretical models, particularly in stars with masses below approximately $\unit[2]{\msun}$ and above $\unit[2.2]{\msun}$. Initially, their rotational velocities, $\langle v/v_\mathrm{ZAMS}\rangle$, exhibit an increase during the early stages of the MS, followed by a gradual decrease as they approach the end of the MS phase. This pattern of acceleration peaks at around $\unit[0.25]{\tms}$ \modified{with a window size of $\unit[0.1]{\tms}$}, as indicated by the first vertical dashed gray line in the figure, a trend consistent across these mass ranges. Interestingly, in the more massive subset ($\langle M\rangle>\unit[2.8]{\msun}$), a second acceleration peak appears at $\unit[0.55]{\tms}$, denoted by the second vertical dashed gray line, albeit this secondary increase is substantially less pronounced in less massive stars. Meanwhile, stars with masses between $\sim\unit[2.0]{\msun}$ and $\sim\unit[2.2]{\msun}$ demonstrate a modest increase in rotational velocity during the first half of their MS lifespan.

This characteristic closely mirrors observations by \citet{2012A&A...537A.120Z}, {where the authors collected $v\sin i$ for a sample of 2014 B6- to F2-type stars and compared the observed evolution of rotational velocities with different theoretical models.} \commented{The analysis in their work is the most relevant. So I feel necessary to discuss the differences in detail.} They reported rapid velocity increases in the initial third of the MS phase, peaking around $\unit[0.3-0.4]{\tms}$ for considered masses of 2, 2.5, and $\unit[3]{\msun}$. However, our findings diverge in several respects: Firstly, the early acceleration observed in stars approximately $\unit[2]{\msun}$ is less pronounced in our analysis than in \citet{2012A&A...537A.120Z}, where peak velocity increased by a factor of around 1.3. In our dataset, such significant increases in rotational velocity exceeding $\langle v/v_\mathrm{ZAMS}\rangle>1.3$ are only seen in stars around $\unit[1.8]{\msun}$, diminishing as mass increases. Notably, even as acceleration becomes apparent again for stars exceeding $\unit[2.35]{\msun}$, it does not surpass the magnitude observed in lower-mass stars. Secondly, while \citet{2012A&A...537A.120Z} identified the maximum $\langle v/v_\mathrm{ZAMS}\rangle$ at $t/\tms$ roughly between 0.3 and 0.4, our analysis suggests a slightly earlier peak at 0.25. Lastly, the potential second acceleration phase around $t/\tms \sim 0.55$, also noted by \citet{2012A&A...537A.120Z}, occurred later in their analysis, within the $0.6 < t/\tms < 0.7$ interval. The last two discrepancies might stem from different binning on age between our study and theirs.

In the latter half of the MS lifespan, the rotational velocity curves are characterized by gradual deceleration, interspersed with minor fluctuations. A notable exception occurs within the mass range of approximately $\unit[2.0]{\msun}$ to $\unit[2.1]{\msun}$, where the most dominant peak in velocity is observed at $t/\tms=0.7$, a feature unique to this mass bin. For stars more massive than $\unit[2.5]{\msun}$, a consistent and monotonical decline in rotation is observed until the MS ends. By analyzing the phase between $t/\tms=0.6$ and $t/\tms=0.9$, we found that the deceleration rates exceed those predicted by the differential rotation model (approximately $-0.6$), reaching a minimum rate of $-1.50\pm0.16$ for stars around $\unit[2.8]{\msun}$. While this linear relationship between $\langle v/v_\mathrm{ZAMS}\rangle$ and age is most pronounced in more massive stars, the deceleration in stars with $M<\unit[2]{\msun}$ during their late MS phase can also be approximated by a gradual slowing. Moreover, the slopes between $t/\tms=0.25$ and $0.55$ for various mass bins (excluding the $2.0 < M/\msun < 2.1$ range) are gentler compared to later stages ($0.6 < t/\tms<0.9$), hinting at the presence of a second acceleration phase concluding at $t=\unit[0.55]{\tms}$.

In every mass bin, we observed a distinctive `hook'-like feature in the rotational velocity curves at the terminal phase of the MS, lasting for a short period of $\Delta t \approx \unit[0.05]{\tms}$. This temporary increase in rotation likely reflects the star's overall contraction when it evolves off the MS, as detailed by \citet[][{their Figure~15}]{2008A&A...478..467E}. This contraction phase is triggered when the hydrogen within the convective core dwindles, no longer sufficient to sustain the star's equilibrium structure. Consequently, the star contracts until hydrogen burning recommences in the surrounding shell, leading to a brief spike in rotational speed.

\subsection{Evolution of rotation's dependence on metallicity}
\label{sec:evolution_metal}

\begin{figure*}[ht!]
\centering
\includegraphics{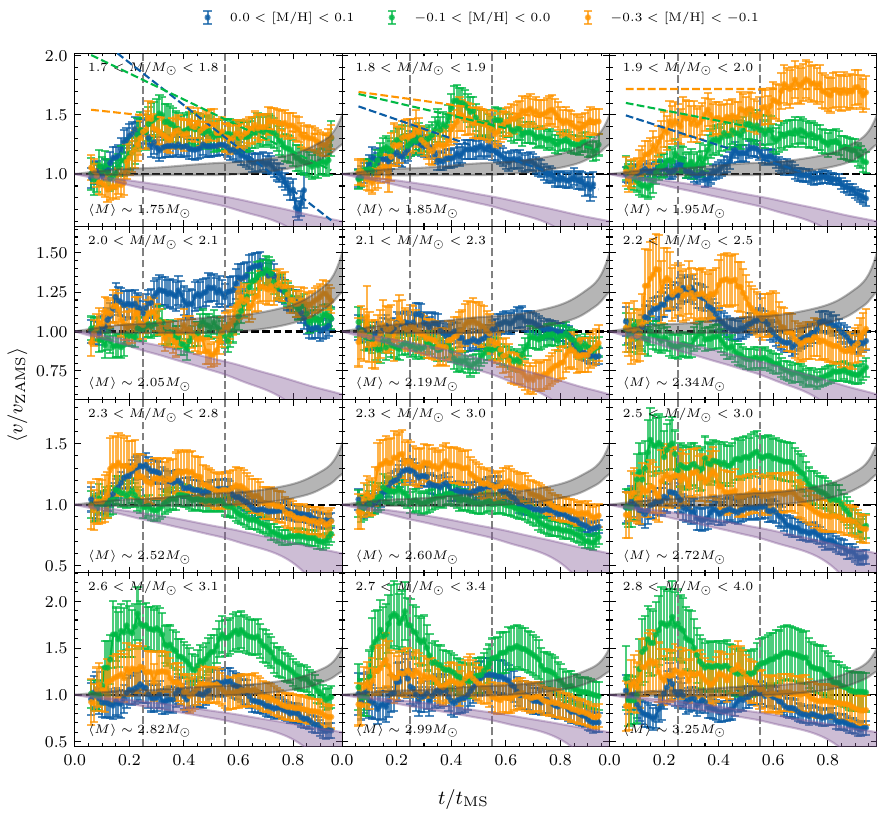}
\caption{Similar to Figure~\ref{fig:vevo_zams_mass}, but further binned by metallicity range: $\unit[0.0]{dex} < \mathrm{[M/H]} < \unit[0.1]{dex}$ (blue), $\unit[-0.1]{dex} < \mathrm{[FeM/H/H]} < \unit[0.0]{dex}$ (green), and $\unit[-0.3]{dex} < \mathrm{[M/H]} < \unit[-0.1]{dex}$ (orange). In the top row, the deceleration phase in the second half of the MS is indicated by the dashed lines with the same color. \label{fig:vevo_zams}}
\end{figure*}

Figure~\ref{fig:vevo_zams_mass} is extended to analyze metallicity's influence on rotational velocity, categorized into three metallicity ranges: two near solar --- $\unit[0.0]{dex} < \mathrm{[M/H]} < \unit[0.1]{dex}$ (blue) and $\unit[-0.1]{dex} < \mathrm{[M/H]} < \unit[0.0]{dex}$ (green) --- and one for metal-poor stars, $\unit[-0.3]{dex} < \mathrm{[M/H]} < \unit[-0.1]{dex}$ (orange). This selection is guided by two main considerations: firstly, ensuring each metallicity bin contains a sufficiently large sample for robust statistical analysis; and secondly, avoiding the super-solar metallicity range where models become less reliable due to increased opacity from heavy elements. These metallicity intervals allow for a comparison across similarly sized samples, although disparities may arise in certain mass bins. The average metallicity values for these bins are $\unit[0.05]{dex}$, $\unit[-0.05]{dex}$, and $\unit[-0.18]{dex}$, respectively.

The observations detailed in Figure~\ref{fig:vevo_zams_mass} largely remain unchanged in Figure~\ref{fig:vevo_zams}, demonstrating that: (1) early acceleration reaching a peak at $t/\tms\sim 0.25$ occurs independently of metallicity within the studied range; (2) beyond the mid-point of the MS lifespan, rotational velocity generally decreases gently, except for the mass bin $2.0 < M/\msun < 2.1$ which deviates from this pattern, and a noticeable shift in the location of the second peak for stars with $M/\msun < 2.7$ across different metallicity ranges; (3) the `hook' feature is consistently present. These consistent observations across our analyses suggest that the identified features are not mere artifacts of sample selection but likely reflect inherent characteristics of A-type stars.

In the first row of Figure~\ref{fig:vevo_zams}, {the deceleration phase in the second half of the MS could be approximated by a linear decline (dashed lines). Based on the slope of this approximation, we notice} a metallicity-dependent deceleration pattern, where the metal-poor subsample exhibits a more gradual slowdown compared to their metal-rich counterparts, with deceleration rates intensifying as average metallicity increases. Specifically, the deceleration slopes for the metal-rich subsample, steeper than those predicted by the differential rotation model, reach their lowest at $-1.80\pm0.24$ for stars in the least massive range ($1.7<M/\msun < 1.8$). This trend of varying slopes with stellar mass is consistent across all metallicity bins, showcasing less steep declines as stellar mass increases. For example, in the intermediate metallicity range ($\unit[-0.1]{dex} < \mathrm{[M/H]} < \unit[0.0]{dex}$), slopes shift from $-1.10\pm0.12$ to a less steep $-0.44\pm0.10$ as stars gets more massive. Remarkably, for the most metal-poor subsample, the curve even becomes almost flat in $1.9 < M/\msun < 2.0$ with a slope of $-0.00\pm 0.13$. It's noted that for metal-rich stars in the mass range $1.7<M/\msun < 1.8$, the evolutionary curve is truncated before the MS's end, primarily due to a temperature cutoff at $\unit[7000]{K}$, which excludes late MS phase data for low-mass, metal-rich stars.

Yet, this distinct metallicity dependence is less pronounced in stars exceeding $\sim\unit[2.0]{\msun}$ in mass, especially within the range of $2$ to $\unit[2.5]{\msun}$. Here, the rotational evolution does not exhibit a consistent pattern across different metallicity levels, as illustrated in the second row of Figure~\ref{fig:vevo_zams}. It's only when the average stellar mass reaches approximately $\unit[2.6]{\msun}$ that more predictable patterns emerge once again. Contrary to expectations, the evolution profile for the slightly metal-poor subgroup, particularly around $\langle M \rangle \sim\unit[2.7]{\msun}$, significantly diverges from the anticipated intermediate behavior between the most metal-poor and metal-rich groups. This slightly metal-poor subgroup experiences notably sharp acceleration in the early MS phase and pronounced deceleration towards its conclusion. This discrepancy might be attributed to stochastic effects, exacerbated by the limited number of young, massive stars in our dataset. This limitation, as highlighted in the top panel of Figure~\ref{fig:prop}, introduces significant uncertainty in the initial velocity estimates ($\langle v_\mathrm{ZAMS}\rangle$), affecting the observed evolutionary trends for $\langle v/v_\mathrm{ZAMS}\rangle$. Such deviations suggest that inaccuracies in determining $\langle v_\mathrm{ZAMS}\rangle$ for the slightly metal-poor subgroup may lead to an underestimation of their true evolutionary dynamics, which, if corrected, would likely position their trajectory between those observed for the metal-rich and metal-poor groups.

\begin{figure*}[ht!]
\centering
\includegraphics{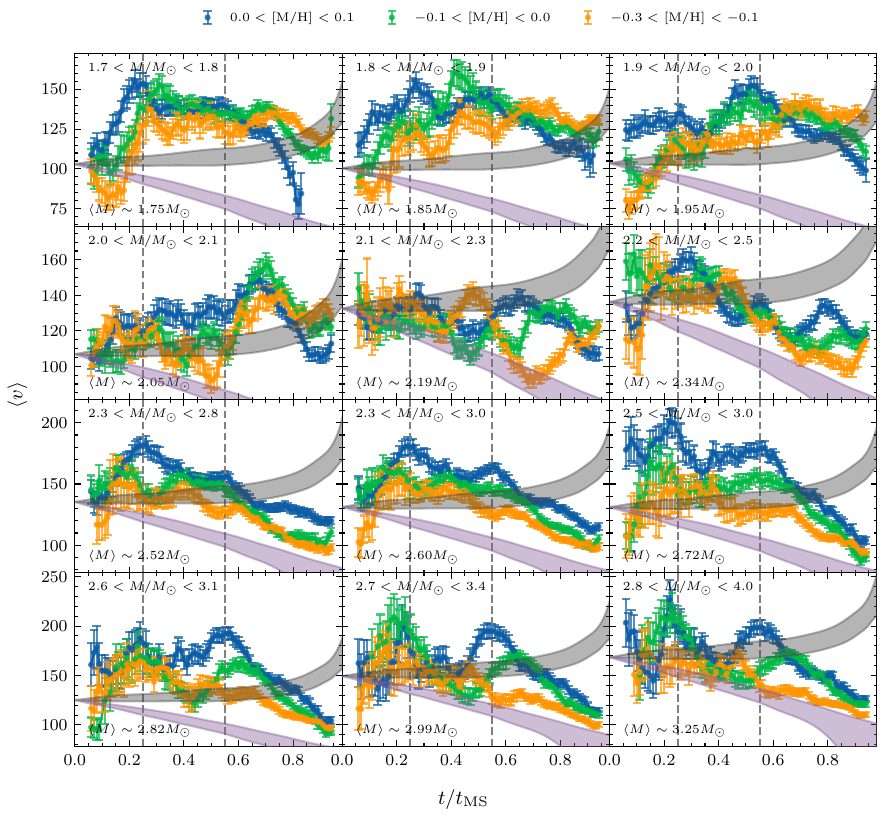}
\caption{Similar to Figure~\ref{fig:vevo_zams}, but for average equatorial velocity $\langle v \rangle$. \modified{In each mass bin, we adopted the average values of $\langle v_\mathrm{ZAMS}\rangle$ across three metallicity bins as the value of $\langle v_\mathrm{ZAMS}\rangle$ for the models.}\label{fig:vevo}}
\end{figure*}

To alleviate the reliance on the $\langle v_\mathrm{ZAMS}\rangle$, we introduce an analogous figure showcasing the average equatorial velocity $\langle v \rangle$ in Figure~\ref{fig:vevo}. This approach synthesizes evolutionary effects with the initial rotational states, presenting evolution profiles that start from actual $\langle v_\mathrm{ZAMS}\rangle$ values rather than a normalized starting point. As a result, this method substantially reduces the associated uncertainties in the profiles. The first row of the figure illustrates deceleration slopes that continue to emphasize metallicity's influence, indicating that the more metal-rich samples may experience greater angular momentum loss on the surface. However, the consistency in evolutionary order observed for less massive stars does not hold for those up to $\unit[2.5]{\msun}$. In contrast, for stars more massive than this threshold, clear metallicity-driven patterns reappear, with the evolution curve for $\unit[-0.1]{dex} < \mathrm{[M/H]} < \unit[0.0]{dex}$ fitting neatly between those of the other two metallicity groups—a distinction not as apparent in Figure~\ref{fig:vevo_zams}. This discrepancy suggests that the scaling in the previous figure might be influenced by the precision of $\langle v_\mathrm{ZAMS}\rangle$ estimates. Intriguingly, for stars averaging more than $\unit[2.7]{\msun}$ in mass, a similar metallicity relationship emerges, where the $\unit[0.0]{dex} < \mathrm{[M/H]} < \unit[0.1]{dex}$ and $\unit[-0.1]{dex} < \mathrm{[M/H]} < \unit[0.0]{dex}$ bins show much steeper slopes than the most metal-poor subgroup. These patterns suggest a metallicity-correlated angular momentum loss rate, a phenomenon that appears to be consistent across different stellar masses, even as deceleration rates vary according to both mass and metallicity.

Now, we move our focus to $\langle v_\mathrm{ZAMS} \rangle$, which represents the initial status of rotation. For the first three mass bins where $\langle M \rangle$ is below $\unit[2]{\msun}$, $\langle v_\mathrm{ZAMS} \rangle$ increases as the metallicity gets larger. When the mass is in the transition domain between $2$ and $\unit[2.5]{\msun}$, it is hard to confirm any difference between these samples, and for more massive subsamples, the conclusion is hindered by the limited sample size at the very point of ZAMS. As mentioned, an underestimation of $\langle v_\mathrm{ZAMS} \rangle$ for $\unit[-0.1]{dex} < \mathrm{[M/H]} < \unit[0.0]{dex}$ could explain the outlier behavior in the last row of Figure~\ref{fig:vevo_zams}. It is worth noticing that, while the ZAMS velocities of various metallicity might be consistent with the errors, generally speaking, metal-rich stars show larger rotational velocity on the MS lifetime compared to slightly metal-poor and metal-poor subsamples for $\langle M \rangle>\unit[2.5]{\msun}$. This is most obvious after the possible second peak in Figure~\ref{fig:vevo}, where the blue evolution track is positioned above the green track, while the green track, in turn, lies above the orange track. Such a feature is different from that of the less massive stars which, initially, the metal-rich stars rotate faster than metal-poor ones, but a remarkable reversal occurs beyond the second peak, with metal-rich stars slowing down more swiftly. This rapid deceleration, associated with higher metallicity, as delineated in Figure~\ref{fig:vevo_zams}, suggests that increased metallicity significantly influences the {surface} angular momentum loss, {either due to angular momentum redistribution or mass loss,} leading to a quicker reduction in rotational speed.

\subsection{Rapid rotators as a function of metallicity}
\label{sec:rapid}
In the previous section, we discussed the evolution of rotation as a function of MS lifetime and their dependence on stellar mass and metallicity. While such an approach can simplify the analysis by focusing on how these specific factors influence stellar rotation, it requires a significant number of stars and thus is only possible for stars around solar metallicity in the Milky Way (MW). To extend our understanding to lower metallicity regimes, we adopt an alternative approach: comparing rotation rates of a given population within the MW to those in the Small Magellanic Cloud (SMC) and Large Magellanic Cloud (LMC) dwarf galaxies \citep[e.g.,][]{1999A&A...346..459M, 2006A&A...452..273M, 2007A&A...462..683M}. These dwarf galaxies enable us to explore metallicity effects down to $\unit[0.14]{\zsun}$ ($\mathrm{[M/H]}\sim \unit[-0.82]{dex}$) for the SMC and $\unit[0.5]{\zsun}$ ($\mathrm{[M/H]}\sim \unit[-0.30]{dex}$) for the LMC. Such comparisons, however, often grapple with the complexities of analyzing mixed populations or cluster members whose formation environment might be unique. These challenges are compounded by typically small sample sizes of less than a thousand, hindering the segregation of stars into distinct bins for a close examination of variations within specific subsets. In our study, we leverage our extensive catalog, categorizing stars by mass, to assess how the proportion of rapid rotators varies with metallicity, extending to $\mathrm{[M/H]}=\unit[-0.65]{dex}$, thereby providing a more detailed picture of rotational behaviors across different metallicity levels.

\begin{figure}[ht!]
\centering
\includegraphics{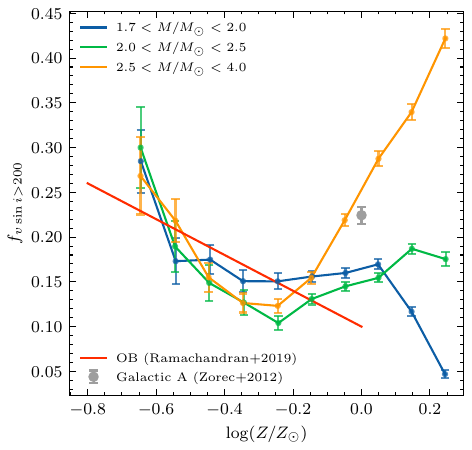}
\caption{Fraction of stars with $v\sin i$ higher than $\unit[200]{km\,s^{-1}}$ as a function of $\log(Z/Z_\odot)$ binned by stellar mass: $1.7 < M/\msun < 2.0$ (blue), $2.0 < M/\msun < 2.5$ (green), and $2.5 < M/\msun < 4.0$ (orange). Galactic A-type star sample from \citet{2012A&A...537A.120Z} is marked as a grey point with assumed solar metallicity. The red line represents the relation found in OB-type stars from \citet{2019A&A...625A.104R}, with SMC data from \citet{2007A&A...472..577M, 2019A&A...625A.104R}, LMC data from \citet{2008A&A...479..541H, 2013A&A...550A.109D,2013A&A...560A..29R, 2018A&A...609A...7R} and MW data from \citet{2006A&A...457..265D, 2012AJ....144..130B, 2014A&A...562A.135S} \commented{The data is to reproduce the result in \citet{2019A&A...625A.104R}, so I just used those included in their work.}.
\label{fig:vsini_metal_trend}}
\end{figure}

We define fast-rotating stars as those with $v\sin i$ exceeding $\unit[200]{km,s^{-1}}$, following the criteria used by \citet{2018A&A...609A...7R} and \citet{2019A&A...625A.104R}. Our analysis divides stars into three mass bins, based on their behavior seen in Figure~\ref{fig:vevo_zams}: low-mass stars ($1.7 < M/\msun < 2.0$, blue) experiencing a metallicity-dependent decrease in rotation rates; intermediate-mass stars ($2.0 < M/\msun < 2.5$, green) in a transitional phase; and high-mass stars ($2.5 < M/\msun < 4.0$, orange) that exhibit a monotonical decrease after the second peak. In Figure~\ref{fig:vsini_metal_trend}, we observe linear relationships between the fraction of fast rotators and metallicity for stars with $\mathrm{[M/H]}<\unit[-0.2]{dex}$, noting substantial variability across mass bins at super-solar metallicity levels. This study follows the methodology of \citet{2019A&A...625A.104R}, who compared OB stars' rotational velocities across the MW, LMC, and SMC. Our findings for metal-poor A-type stars align closely with these OB stars' rotational trends. Interestingly, A-type stars with solar metallicity exhibit a higher fraction of rapid rotators than their OB counterparts in the MW. This aligns with \citet{2012A&A...537A.120Z} and is supported by the velocity trends across spectral types described by \citet{1986PASP...98.1233S} that field and cluster stars have mean projected rotational velocity increasing from below $\unit[20]{km\,s^{-1}}$ at spectral type G0 to around $\unit[200]{km\,s^{-1}}$ at spectral type A, and then gently decreasing to $\unit[150]{km\,s^{-1}}$ at spectral type earlier than B0. Additionally, the observed correlation between lower metallicity and higher rotational speeds among OB stars, potentially with a steeper slope for A-type stars, suggests a universal pattern in stellar rotational velocity evolution across different metallic environments. It's noteworthy that while variations among A-type stars' mass bins are minimal, the dataset for OB stars exhibits greater variability \citep[see their Figure~8]{2019A&A...625A.104R}, highlighting the diverse dynamics of stellar rotation in different galactic contexts.

\section{Discussion}
\label{sec:discussion}
\subsection{Mechanisms controlling the evolution of rotation}
\label{sec:mechanism}

Upon settling on the MS with a given angular momentum, a star's rotational evolution is predominantly influenced by a trio of physical processes\footnote{\modified{Interaction with a companion by tides or mergers could affect surface rotation as well, e.g., the orbital decay of a planet due to tidal forces could transform angular momentum to the red giant stars \citep{2016A&A...591A..45P}.}}:
\begin{itemize}
	\item local conservation of the angular momentum
	\item internal transport mechanisms
	\item stellar winds
\end{itemize}
Local angular momentum conservation leads to adjustments in rotational velocity as the star undergoes radial expansion. Internal transport mechanisms can redistribute the angular momentum throughout the star through meridional circulation \citep{1992A&A...265..115Z} and the Gratton–\"{O}pik cell at the outer envelope carries angular momentum from the inner part of the star to the surface, tending to accelerate it \citep{2009pfer.book.....M}. As for stellar winds, they can remove angular momentum from the stellar surface through the mass loss process.  In contrast, stellar winds can act to decelerate the star by stripping angular momentum from the surface through mass loss. The interplay between these processes---whereby the first and last tend to decelerate surface rotation, counterbalanced by the acceleration due to internal transport---dictates the trajectory of a star's rotational velocity on the MS.

While stellar winds are pivotal in the angular momentum evolution of massive stars, affecting their ultimate fate \citep{2008A&ARv..16..209P}, they are anticipated to have a negligible impact on the angular momentum evolution of A-type stars considered in this study. Observations of the H$\alpha$ line profile in A-type MS stars suggest an upper mass loss limit of up to $1$ and $\unit[2\times10^{-10}]{\msun,yr^{-1}}$ \citep{1992A&A...257..663L}, and typically in the range of $10^{-12}$ to $\unit[10^{-10}]{\msun,yr^{-1}}$ \citep{1999isw..book.....L}. Even for stars rotating near their critical velocity, the rotationally induced enhancement in mass loss remains modest, less than a fivefold increase, not exceeding $\unit[10^{-11}]{\msun,yr^{-1}}$ \citep{2022A&A...665A.126N}. Therefore, the principal drivers of rotational evolution in this context are local angular momentum conservation and internal transport mechanisms, not stellar winds. Models from \citet{2012A&A...537A.146E, 2022A&A...665A.126N} indicate a balance between inefficient internal transport and minimal mass loss, leading stars to evolve with nearly constant rotational velocities for masses up to $\unit[10]{\msun}$. However, as seen in Figure~\ref{fig:vevo_zams_mass}, the observed acceleration in the early MS phase exceeds predictions from both solid body and differential rotation scenarios. This contradicts findings of \citet{2007AJ....133.1092W} that stars in the $\unit[6-12]{\msun}$ range, show no significant rotational speed variation through the MS from $t\sim 1$ and $\unit[15]{Myr}$, challenging the expectation of a uniform rotational behavior across different stellar mass ranges.

Since the gradual expansion of stellar radii would inherently slow down the surface rotational velocity, any observed acceleration in rotation at early phases of the MS must stem from internal redistribution of angular momentum, barring external influences like close binary interactions. The amplitude of the meridional circulation's radial component undergoes a sign change near the stellar surface, enabling the transport of angular momentum from the star's inner regions to its surface \citep{1945MmSAI..17....5G, 1951MNRAS.111..278O}. Although this mechanism is theoretically capable of increasing surface rotation \citep[see their Figure~18]{2008A&A...478..467E}, such acceleration typically occurs later in the MS phase ($t/\tms > 0.6$), difficult to reconcile with the early and pronounced speed-up depicted in Figure~\ref{fig:vevo_zams_mass}.

Another aspect to consider is the initial adjustment phase which depends on the model's assumed initial rotation. A brief period of adjustment occurs at the start, as shear turbulence becomes active, eroding gradients established by meridional circulation and leading the rotation rate ($\Omega$) profile toward an equilibrium configuration \citep{1992A&A...265..115Z, 1999A&A...341..181D}. Since models typically assume uniform rotation at the ZAMS, meridional circulation shifts angular momentum from the star's outer regions to its core, noticeably decelerating the surface rotation for a tiny portion of the MS lifetime \citep{2008A&A...478..467E, 2022A&A...665A.126N}. This assumption of initial rigid rotation, to minimize the total angular momentum constrained by the ratio of rotational kinetic energy to the gravitational potential, finds justification in (a) dynamic stability against axisymmetric perturbations, advocated for solid-body stellar rotators \citep{1987A&A...176...53F}, and (b) the promotion of angular momentum redistribution by turbulent viscosity \citep{2008A&A...479L..37M}. Thus, it's generally believed that stars exhibit uniform rotation throughout the fully convective pre-MS stage. Despite its widespread application, direct evidence for this assumption remains elusive.


The relaxation timescale from initial conditions to a steady state in stellar rotation shows a significant dependence on surface rotational velocity, as delineated by \citet{1999A&A...341..181D}. The formula 
$$\tau_\mathrm{rel}\propto\tau_\mathrm{KH}{\Omega_\mathrm{S}}^{-2}\left(\frac{M^2}{R^3}\right)$$
reveals that $\tau_{\mathrm{rel}}$, the relaxation timescale, scales with the Kelvin-Helmholtz timescale $\tau_{\mathrm{KH}}$, where $\Omega_{\mathrm{S}}$ represents the angular velocity at the star's surface. This relationship suggests that for a differentially rotating star at the ZAMS with very low angular velocities, reaching a steady state could extend significantly, potentially matching the star's lifespan. For stars initiating rotation at high velocities ($v_{\mathrm{s}} > \unit[100]{km\,s^{-1}}$), the relaxation process is too short to be observable. Conversely, for stars beginning with rotational velocities around $\unit[10]{km\,s^{-1}}$, permissible under differential rotation models without minimizing total angular momentum, this could imply a prolonged adjustment period and observable surface acceleration in the early MS phase. This might challenge the assumption of uniform rotation at ZAMS, particularly for intermediate-mass A-type stars, suggesting a slower initial surface rotation. This aligns with findings by \citet{2004ApJ...601..979W} that stars do not maintain solid-body rotation during their transition from convective to radiative phases in the Orion star-forming region, encompassing stars ranging from $0.4$ to over $\unit[10]{\msun}$.

The tentative second peak in rotational velocity observed around $t/\tms\sim 0.55$ is hard to explain for the relaxation, as a steady-state rotation profile would presumably have been reached after the initial peak. This suggests that any significant readjustment phase at this later stage, especially one that results in acceleration, would necessitate some form of internal angular momentum shift not accounted for in the initial relaxation. Given that the surface rotation has already been accelerated, the duration of this secondary acceleration phase should be shorter than that of the initial peak. Therefore, this later acceleration could be linked to the Gratton–\"{O}pik term, indicating a more intense than expected deepening of the outer zone where inverse circulation occurs during MS evolution \citep{2000A&A...361..101M}. Such a scenario hints at the absence of a truly stationary state in stellar rotation during the MS phase, challenging the simplifications used in some stellar evolution models.

In the latter stages of the MS, the evolutionary path of stars is generally characterized by a monotonical deceleration, with the notable exception of those in the mass range $2.0 < M/\msun < 2.1$. This observation suggests that the dominant factor influencing this phase is the loss of angular momentum at the stellar surface, attributable more to the star's radial expansion than to stellar wind-induced mass loss. The observed steepening of the deceleration slope with increasing stellar mass, although potentially exceeding theoretical predictions, can be possibly reconciled with the model for two main reasons: Firstly, the measured value of $\langle v/v_\mathrm{ZAMS}\rangle$ --- and consequently, its slope --- could be inaccurately determined due to erroneous estimations of $\langle v_\mathrm{ZAMS}\rangle$, particularly for higher mass stars. Secondly, the differential rotation model employed by \citet{2012A&A...537A.120Z} is predicated on solar metallicity assumptions. In contrast, our study imposes a metallicity range from $-0.3$ to $\unit[0.1]{dex}$, with an average slightly below solar ($\unit[-0.07]{dex}$). This discrepancy in metallicity could account for the observed variation in slope.

\subsection{Abnormity between $\unit[2.0]{\msun}$ and $\unit[2.1]{\msun}$}
\label{sec:transition}
Given the complexity of stellar evolution and rotational dynamics, the phenomena observed in the mass range of $2.0$ to $\unit[2.3]{\msun}$ stand out as particularly intriguing. Within this mass range, a transition occurs from a unimodal to a bimodal distribution of rotational velocities, as identified by \citetalias{2021ApJ...921..145S}. A notable deviation in rotational velocity is observed at $2.0 < M/\msun < 2.1$, where the anticipated continuous decline in rotational velocity following the first or second peak is disrupted by a significant peak at $t/\tms \sim 0.7$, a feature absent in other mass bins. This anomaly persists across all three examined metallicity bins and does not vary with the choice of $\langle v/v_\mathrm{ZAMS}\rangle$ or $\langle v \rangle$, suggesting it could be an intrinsic characteristic or result from systematic bias within this specific subset.

If the emergence of this delayed peak results from enhanced angular momentum transfer to the surface due to inverse circulation, it challenges explanations for its delayed appearance by approximately $\delta t\sim 0.15$ of the MS lifetime. Furthermore, the magnitude of this acceleration does not align with the second peak observed around $t/\tms\sim0.55$ in other mass ranges. The absence of this delayed peak in prior studies casts doubt on its authenticity, suggesting the influence of a bias unique to certain data subsets. One potential source of bias could relate to the effective temperature determination around $\unit[9500]{K}$ \citep{2022A&A...662A..66X}, where the spectral gradients of hydrogen lines in A0 stars diminish, potentially leading to skewed interpretations for stars with masses between $\unit[2.0]{\msun}$ and $\unit[2.2]{\msun}$. Additional investigations are warranted to clarify the origins and implications of this anomalous behavior, highlighting the need for careful consideration of potential biases in stellar rotational studies.

\subsection{Influence of metallicity}
\label{sec:metal}
The role of metallicity in shaping the rotational characteristics of stars, particularly throughout the MS phase, is intricately linked to three fundamental mechanisms outlined in Section~\ref{sec:mechanism}:
\begin{itemize}
	\item Under similar distributions of initial angular momentum, stars having lower metallicity rotate faster because they undergo less violent nuclear reactions, making their radii at the ZAMS smaller than those of metal-rich stars, {extending down to the metallicity levels observed in the Magellanic Clouds} \citep{1999A&A...346..459M}. However, the subsequent rapid expansion of metal-poor stars leads to a slightly more pronounced decrease in rotational velocity over time.
	\item The Gratton–\"{O}pik cell is proportional to $1/\rho$, which gets more prounced at the outer envelop and lower $Z$. This makes the transfer of angular momentum by this inverse meridional circulation more efficient with a decreasing metallicity \citep[{from $Z=0.02$ to $Z=0.002$, or even lower, }][]{2008A&A...478..467E, 2009IAUS..256..337H}.
	\item The mass loss due to radiation-driven stellar winds is linked with the metallicity of the star, as higher metallicity increases the number of lines available for the absorption and scattering of radiation, thus driving the stellar outflow more effectively. This relationship between mass-loss rate and metallicity follows a power-law distribution of line strengths, as detailed by \citet{2001A&A...369..574V}. In contrast, \citet{2006isna.confE..15M} highlighted the relevance of mechanical winds in {very metal-poor ($Z\leqslant \unit[10^{-5}]{\zsun}$)} environments, where the efficacy of radiation-driven winds diminishes. Moreover, rotational mixing can lead to enhanced opacity in the outer layers of a star, potentially reinvigorating line-driven winds. This dynamic interplay between different types of stellar winds and metallicity underscores the complex mechanisms governing stellar mass loss and its consequent impact on stellar rotation and evolution.
\end{itemize}

For the range of stellar masses analyzed here, the effect of stellar wind on surface rotation velocity is minimal. Even when accounting for mechanical winds, \citet{2022A&A...665A.126N} reported a maximum mass loss rate below $\unit[10^{-11}]{\msun\,yr^{-1}}$ across the MS for stars lighter than $\unit[3]{\msun}$. Hence, the variations in rotation velocity as a function of metallicity, as observed in Figures~\ref{fig:vevo_zams} and \ref{fig:vevo}, are likely not due to mass loss but rather to internal angular momentum redistribution. A discernible pattern emerges, where metal-rich samples exhibit a faster decline in both $\langle v/v_\mathrm{ZAMS}\rangle$ and $\langle v\rangle$ for masses below $\unit[2.0]{\msun}$. While this correlation fades in the transitional mass range between $\unit[2.0]{\msun}$ and $\unit[2.3]{\msun}$, it reappears, albeit tentatively, for heavier stars, particularly in terms of $\langle v\rangle$. This behavior aligns with predictions related to meridional circulation, whereby the Gratton–\"{O}pik cell, more effective at lower metallicities, mitigates the reduction in rotational velocity caused by stellar expansion. The trend between the deceleration phase's slope and metallicity aligns with the predictions made by \citet{2013A&A...553A..24G, 2022A&A...665A.126N}, though a direct comparison is challenging due to our analysis involving mixed subpopulations with varied initial rotation velocities.

In Figure~\ref{fig:vevo}, a noticeable discrepancy in $\langle v_\mathrm{ZAMS}\rangle$ across different metallicity groups was observed for stars with lower masses. This trend, where metal-rich populations exhibit higher initial rotational velocities, contradicts the expectation that metal-poor stars should rotate faster due to their comparatively smaller sizes at the ZAMS. It is important to recognize that this assumption presumes the same angular momentum across stars of varying metallicity at the onset of the MS. The discrepancy observed could imply a more efficient process of angular momentum dispersion during the stellar formation phase for metal-poor stars, potentially influenced by stronger fossil magnetic fields, which are theorized to be more intense in stars with lower metallicity \citep{2017A&A...597A..71S}. Should this be verified, it would suggest a more significant role for magnetic braking in the evolution of metal-poor star rotation velocities.

\subsection{Caveats}
\label{sec:caveat}

The analysis of rotation within this study primarily relies on the average values of $\langle v\sin i\rangle$ or $\langle v \rangle$. This approach, while useful for generalizing rotational trends, introduces potential caveats. 

Firstly, the extensive size of the sample in this study presents significant challenges in removing contaminants such as close binaries and chemically peculiar stars. Although efforts were made to identify these objects through radial velocity variations and cross-referencing with existing catalogs, the process is inherently incomplete. The complexity and diversity of stellar phenomena mean that some contaminants inevitably remain, potentially influencing the overall analysis. A more detailed and refined approach to identifying and excluding binary systems and chemically peculiar stars could further enhance the reliability of the findings. This aspect underscores the need for caution in interpreting the results and highlights the potential for future work to refine and corroborate these observations \citep{2023Galax..11...54Z}.

Secondly, our model primarily considers surface rotational velocity as a function of stellar mass, age, metallicity, and initial rotation velocity, while potentially overlooking other influential factors. Conspicuously, magnetic fields could significantly impact internal stellar rotation, structure, and evolution \citep{2022MNRAS.517.2028K}. Magnetic activity, stemming from dynamo-driven fields in stars with convective envelopes \citep{1937MNRAS..97..458F, 1999A&A...349..189S} or fossil fields in early-type stars \citep{2012MNRAS.419..959O}, is known to affect rotational dynamics. Although the detection is quite few, magnetic fields are detected in 10\% of the early-type stars \citep{2012MNRAS.419..959O}, which typically rotate slower than their non-magnetic counterparts due to angular momentum loss via magnetically confined winds \citep{1967ApJ...148..217W, 2009MNRAS.392.1022U}. This relationship between magnetic fields and rotational behavior, supported by studies of rotational modulation of eight CP stars observed by Transiting Exoplanet Survey Satellite \citep{2022MNRAS.517.5340K}, suggests magnetic braking as a critical factor in stellar rotation. The limited detection of magnetic fields in early-type stars, however, underscores the need for further investigation into the broader implications of magnetic activity on stellar rotation across different stellar types and evolutionary stages.

Finally, employing the average values of $v\sin i$ or $v$ to characterize a population's rotational behavior may complicate direct comparisons with models, particularly if the underlying distribution is not singularly peaked. \commented{I think this caveat is about how to interpret the average value of $v\sin i$, so I added a note at the end of the first paragraph in Section~\ref{sec:result}, instead of in Section~\ref{sec:data} where I measured $v\sin i$.} For instance, bimodal distributions of $v\sin i$ have been observed in certain star clusters \citep{2019ApJ...883..182S, 2020MNRAS.492.2177K, 2023MNRAS.518.1505K}, highlighting the potential complexity of stellar rotational velocities. Such distributions, despite converging towards a unified mean value in our analysis, necessitate sophisticated deconvolution methods for accurate delineation, akin to approaches utilized in \citet{2012A&A...537A.120Z} or \citetalias{2021ApJ...921..145S}. However, it's crucial to acknowledge that average rotational velocities remain valid indicators for populations exhibiting multi-peak distributions. Additionally, the phenomenon of split MSs observed in young star clusters \citep{2015MNRAS.450.3750M, 2015MNRAS.453.2637D, 2016MNRAS.458.4368M, 2017MNRAS.465.4363M, 2017ApJ...844..119L,2017MNRAS.467.3628C, 2018MNRAS.477.2640M}, which might suggest divergent rotational speeds, does not significantly affect our age determinations. The bimodal MS seen in clear in the color-magnitude diagram \citep[e.g.,][]{2019ApJ...883..182S} is predominantly attributed to gravity-darkening effects and does not distinctly manifest within the HR diagram when effective temperatures are mainly determined through normalized H$\alpha$ line profiles. Consequently, this bifurcation will not manifest as two branches in the HR diagram, thus not misleadingly categorized as another population with fast rotation at $t/t_\mathrm{MS} \sim 0.1$. 

\section{Summary}
\label{sec:summary}

In this work, we built upon the foundational studies of \citetalias{2021ApJS..257...22S} and \citetalias{2021ApJ...921..145S}, assembling a comprehensive catalog of 104,572 A-type stars, derived from LAMOST MRS DR 9 and spanning effective temperatures between $\unit[7000]{K}$ and $\unit[14500]{K}$. Our analysis focused on a subset of 84,683 `normal' stars—those classified as single, non-CP, non-variable, and not belonging to star clusters. The major insights gained from our study are summarized as follows.

\begin{itemize}
	\item By examining the evolution of surface rotational velocity, categorized by stellar mass, we observed diverse evolutionary profiles. The velocities were normalized to their initial values at the ZAMS, and ages were normalized to the total MS lifetime, allowing a consistent comparison across different stellar masses. The majority of these profiles exhibited a common pattern: an initial rapid acceleration up to $t/\tms = 0.25\pm0.1$, a possible second acceleration peak around $t/\tms = 0.55\pm0.1$ for stars with $\langle M \rangle > \unit[2.5]{\msun}$, followed by a gradual decline, and a distinctive `hook' feature towards the end of the MS phase.
	\item The emergence of the initial acceleration peak surpasses the expectations derived from two extreme theoretical models: the rigid body rotator and the differential rotator. This phenomenon is ascribed to the relaxation phase that commences immediately after the ZAMS. Contrary to the flat angular velocity profile typically assumed in models, a more concentrated profile could result in a significant speed increase. This acceleration phase may persist longer before settling into a steady state.
	\item Stars within the mass range of $2.0<M/\msun<2.1$ exhibit anomalous acceleration peaks at $t/\tms = 0.7$. This behavior does not align neatly with theories of internal angular momentum redistribution. Should external factors not account for this phenomenon, it could suggest an artifact stemming from inaccuracies in the effective temperature ($T_\mathrm{eff}$) estimations around $\unit[9500]{K}$.
	\item The monotonical decline during the second half of MS is regulated by the conservation of angular momentum due to an expansion of stellar radii, accompanied by a significant contribution from inverse meridional circulation. The slope of the deceleration is larger at more massive stars, which manifests a more pronounced spin-down trend compared to the prediction of a completely differential model with no internal angular momentum distribution or external angular momentum loss. 
	\item We categorized our subsample based on metallicity into three bins {around solar}: $\unit[0.0]{dex} < \mathrm{[M/H]} < \unit[0.1]{dex}$, $\unit[-0.1]{dex} < \mathrm{[M/H]} < \unit[0.0]{dex}$, and $\unit[-0.3]{dex} < \mathrm{[M/H]} < \unit[-0.1]{dex}$, enabling an analysis of metallicity's impact. The deceleration phase's slope correlates with metallicity, showing that metal-poor {($\unit[-0.3]{dex} < \mathrm{[M/H]} < \unit[-0.1]{dex}$)} sample experiences a steeper spin-down. This observation can be linked to the Gratton–\"{O}pik cell's increased efficiency at lower densities and metallicity. As this mechanism transports more angular momentum from the star's interior to its surface, it moderates the surface rotational velocity's decline, resulting in a gentler slope for metal-poor samples compared to their metal-rich counterparts.
	\item The metal-poor {($\unit[-0.1]{dex} < \mathrm{[M/H]} < \unit[0.0]{dex}$ and $\unit[-0.3]{dex} < \mathrm{[M/H]} < \unit[-0.1]{dex}$)} subsamples begin with lower rotational velocities at the ZAMS, possibly indicating a metallicity-dependent mechanism for angular momentum removal during the formation of MS stars. This observation suggests that the production of magnetic fields, which varies with metallicity, could play a part in this process.
	\item We observed that the proportion of fast rotators ($v\sin i > \unit[200]{km\,s^{-1}}$) declines with increasing metallicity, up to $\log(Z/Z_\odot)\sim -0.2$. This pattern aligns with findings related to OB-type stars in the SMC and LMC \citep{2006A&A...452..273M, 2007A&A...462..683M, 2019A&A...625A.104R}. However, this correlation diverges at and above solar metallicity, suggesting a difference in stellar mass{, or it could be attributed to uncertainties arising from mass loss, which become increasingly significant at higher metallicities}.
\end{itemize}

\begin{acknowledgements}
We thank the anonymous referee for their valuable comments. The Guoshoujing Telescope (the Large Sky Area Multi-Object Fiber Spectroscopic Telescope; LAMOST) is a National Major Scientific Project built by the Chinese Academy of Sciences. Funding for the project has been provided by the National Development and Reform Commission. LAMOST is operated and managed by the National Astronomical Observatories, Chinese Academy of Sciences. This work has made use of data from the European Space Agency (ESA) mission \textit{Gaia} (\url{https://www.cosmos.esa.int/gaia}), processed by the \textit{Gaia} Data Processing and Analysis Consortium (DPAC; \url{https://www.cosmos.esa.int/web/gaia/dpac/consortium}). Funding for the DPAC has been provided by national institutions, in particular the institutions participating in the \textit{Gaia} Multilateral Agreement.

\newline
{\it Facility:} LAMOST
\newline
{\it Software:} PARSEC \citep[2.0;][]{2022A&A...665A.126N}, astropy \citep{2013A&A...558A..33A}, IPython \citep{2007CSE.....9c..21P}, SPInS \citep{2020A&A...642A..88L}, laspec \citep{2021ApJS..256...14Z}, SLAM \citep{2020ApJS..246....9Z}, matplotlib \citep{2007CSE.....9...90H}
\end{acknowledgements}


\begin{thebibliography}{}
\bibitem[Abt(1981)]{1981ApJS...45..437A}Abt, H.~A.\ 1981, \apjs, 45, 437
\bibitem[Abt \& Morrell(1995)]{1995ApJS...99..135A}Abt, H.~A. \& Morrell, N.~I.\ 1995, \apjs, 99, 135
\bibitem[Aerts et al.(2019)]{2019ARA&A..57...35A}Aerts, C., Mathis, S., \& Rogers, T.~M.\ 2019, \araa, 57, 35
\bibitem[Ando(1980)]{1980Ap&SS..73..159A}Ando, H.\ 1980, \apss, 73, 159
\bibitem[Astropy Collaboration et al.(2013)]{2013A&A...558A..33A}Astropy Collaboration, Robitaille, T.~P., Tollerud, E.~J., et al.\ 2013, \aap, 558, A33
\bibitem[Bailer-Jones et al.(2021)]{2021AJ....161..147B}Bailer-Jones, C.~A.~L., Rybizki, J., Fouesneau, M., et al.\ 2021, \aj, 161, 147
\bibitem[Bragan{\c{c}}a et al.(2012)]{2012AJ....144..130B}Bragan{\c{c}}a, G.~A., Daflon, S., Cunha, K., et al.\ 2012, \aj, 144, 130
\bibitem[Buldgen et al.(2015)]{2015A&A...583A..62B}Buldgen, G., Reese, D.~R., \& Dupret, M.~A.\ 2015, \aap, 583, A62
\bibitem[B{\'e}trisey et al.(2023)]{2023A&A...673L..11B}B{\'e}trisey, J., Eggenberger, P., Buldgen, G., et al.\ 2023, \aap, 673, L11
\bibitem[Castelli(2005)]{2005MSAIS...8...25C}Castelli, F.\ 2005, Memorie della Societa Astronomica Italiana Supplementi, 8, 25
\bibitem[Chandrasekhar \& M{\"u}nch(1950)]{1950ApJ...111..142C}Chandrasekhar, S. \& M{\"u}nch, G.\ 1950, \apj, 111, 142
\bibitem[Charbonnel \& Lagarde(2010)]{2010A&A...522A..10C}Charbonnel, C. \& Lagarde, N.\ 2010, \aap, 522, A10
\bibitem[Chen et al.(2020)]{2020ApJS..249...18C}Chen, X., Wang, S., Deng, L., et al.\ 2020, \apjs, 249, 18
\bibitem[Correnti et al.(2017)]{2017MNRAS.467.3628C}Correnti, M., Goudfrooij, P., Bellini, A., et al.\ 2017, \mnras, 467, 3628
\bibitem[Cui et al.(2012)]{2012RAA....12.1197C}Cui, X.-Q., Zhao, Y.-H., Chu, Y.-Q., et al.\ 2012, Research in Astronomy and Astrophysics, 12, 1197
\bibitem[D'Antona et al.(2015)]{2015MNRAS.453.2637D}D'Antona, F., Di Criscienzo, M., Decressin, T., et al.\ 2015, \mnras, 453, 2637
\bibitem[de Jong et al.(2019)]{2019Msngr.175....3D}de Jong, R.~S., Agertz, O., Berbel, A.~A., et al.\ 2019, The Messenger, 175, 3
\bibitem[de Mink et al.(2013)]{2013ApJ...764..166D}de Mink, S.~E., Langer, N., Izzard, R.~G., et al.\ 2013, \apj, 764, 166
\bibitem[Denissenkov et al.(1999)]{1999A&A...341..181D}Denissenkov, P.~A., Ivanova, N.~S., \& Weiss, A.\ 1999, \aap, 341, 181
\bibitem[Deutsch(1970)]{1970stro.coll..207D}Deutsch, A.~J.\ 1970, IAU Colloq. 4: Stellar Rotation, 207
\bibitem[Dufton et al.(2013)]{2013A&A...550A.109D}Dufton, P.~L., Langer, N., Dunstall, P.~R., et al.\ 2013, \aap, 550, A109
\bibitem[Dufton et al.(2006)]{2006A&A...457..265D}Dufton, P.~L., Smartt, S.~J., Lee, J.~K., et al.\ 2006, \aap, 457, 265
\bibitem[Ekstr{\"o}m et al.(2012)]{2012A&A...537A.146E}Ekstr{\"o}m, S., Georgy, C., Eggenberger, P., et al.\ 2012, \aap, 537, A146
\bibitem[Ekstr{\"o}m et al.(2008)]{2008A&A...478..467E}Ekstr{\"o}m, S., Meynet, G., Maeder, A., et al.\ 2008, \aap, 478, 467
\bibitem[Ferraro(1937)]{1937MNRAS..97..458F}Ferraro, V.~C.~A.\ 1937, \mnras, 97, 458
\bibitem[Fr{\'e}mat et al.(2023)]{2023A&A...674A...8F}Fr{\'e}mat, Y., Royer, F., Marchal, O., et al.\ 2023, \aap, 674, A8
\bibitem[Fujimoto(1987)]{1987A&A...176...53F}Fujimoto, M.~Y.\ 1987, \aap, 176, 53
\bibitem[Gaia Collaboration et al.(2016)]{2016A&A...595A...1G}Gaia Collaboration, Prusti, T., de Bruijne, J.~H.~J., et al.\ 2016, \aap, 595, A1
\bibitem[Gaia Collaboration et al.(2023)]{2023A&A...674A...1G}Gaia Collaboration, Vallenari, A., Brown, A.~G.~A., et al.\ 2023, \aap, 674, A1
\bibitem[Georgy et al.(2013)]{2013A&A...553A..24G}Georgy, C., Ekstr{\"o}m, S., Granada, A., et al.\ 2013, \aap, 553, A24
\bibitem[Gratton(1945)]{1945MmSAI..17....5G}Gratton, L.\ 1945, \memsai, 17, 5
\bibitem[Haemmerl{\'e} et al.(2017)]{2017A&A...602A..17H}Haemmerl{\'e}, L., Eggenberger, P., Meynet, G., et al.\ 2017, \aap, 602, A17
\bibitem[Hastings et al.(2021)]{2021A&A...653A.144H}Hastings, B., Langer, N., Wang, C., et al.\ 2021, \aap, 653, A144
\bibitem[Heger \& Langer(2000)]{2000ApJ...544.1016H}Heger, A. \& Langer, N.\ 2000, \apj, 544, 1016
\bibitem[Hirschi et al.(2009)]{2009IAUS..256..337H}Hirschi, R., Ekstr{\"o}m, S., Georgy, C., et al.\ 2009, The Magellanic System: Stars, Gas, and Galaxies, 256, 337
\bibitem[Howarth \& Smith(2001)]{2001MNRAS.327..353H}Howarth, I.~D. \& Smith, K.~C.\ 2001, \mnras, 327, 353
\bibitem[Huang \& Gies(2006)]{2006ApJ...648..580H}Huang, W. \& Gies, D.~R.\ 2006, \apj, 648, 580
\bibitem[Hunt \& Reffert(2023)]{2023A&A...673A.114H}Hunt, E.~L. \& Reffert, S.\ 2023, \aap, 673, A114
\bibitem[Hunter et al.(2008)]{2008A&A...479..541H}Hunter, I., Lennon, D.~J., Dufton, P.~L., et al.\ 2008, \aap, 479, 541
\bibitem[Hunter(2007)]{2007CSE.....9...90H}Hunter, J.~D.\ 2007, Computing in Science and Engineering, 9, 90
\bibitem[H{\"u}mmerich et al.(2020)]{2020A&A...640A..40H}H{\"u}mmerich, S., Paunzen, E., \& Bernhard, K.\ 2020, \aap, 640, A40
\bibitem[Kamann et al.(2020)]{2020MNRAS.492.2177K}Kamann, S., Bastian, N., Gossage, S., et al.\ 2020, \mnras, 492, 2177
\bibitem[Kamann et al.(2023)]{2023MNRAS.518.1505K}Kamann, S., Saracino, S., Bastian, N., et al.\ 2023, \mnras, 518, 1505
\bibitem[Keszthelyi et al.(2022)]{2022MNRAS.517.2028K}Keszthelyi, Z., de Koter, A., G{\"o}tberg, Y., et al.\ 2022, \mnras, 517, 2028
\bibitem[Kobzar et al.(2022)]{2022MNRAS.517.5340K}Kobzar, O., Khalack, V., Bohlender, D., et al.\ 2022, \mnras, 517, 5340
\bibitem[Kroupa(2001)]{2001MNRAS.322..231K}Kroupa, P.\ 2001, \mnras, 322, 231
\bibitem[Kroupa et al.(2013)]{2013pss5.book..115K}Kroupa, P., Weidner, C., Pflamm-Altenburg, J., et al.\ 2013, Planets, Stars and Stellar Systems. Volume 5: Galactic Structure and Stellar Populations, 115
\bibitem[Lamers \& Cassinelli(1999)]{1999isw..book.....L}Lamers, H.~J.~G.~L.~M. \& Cassinelli, J.~P.\ 1999, Introduction to Stellar Winds, by Henny J. G. L. M. Lamers and Joseph P. Cassinelli, pp. 452. ISBN 0521593980. Cambridge, UK: Cambridge University Press, June 1999., 452
\bibitem[Lanz \& Catala(1992)]{1992A&A...257..663L}Lanz, T. \& Catala, C.\ 1992, \aap, 257, 663
\bibitem[Lebreton \& Reese(2020)]{2020A&A...642A..88L}Lebreton, Y. \& Reese, D.~R.\ 2020, \aap, 642, A88
\bibitem[Li et al.(2017)]{2017ApJ...844..119L}Li, C., de Grijs, R., Deng, L., et al.\ 2017, \apj, 844, 119
\bibitem[Li et al.(2019)]{2019MNRAS.487..782L}Li, G., Van Reeth, T., Bedding, T.~R., et al.\ 2019, \mnras, 487, 782
\bibitem[Limongi \& Chieffi(2018)]{2018ApJS..237...13L}Limongi, M. \& Chieffi, A.\ 2018, \apjs, 237, 13.
\bibitem[Liu et al.(2015)]{2015RAA....15.1137L}Liu, C., Cui, W.-Y., Zhang, B., et al.\ 2015, Research in Astronomy and Astrophysics, 15, 1137
\bibitem[Liu et al.(2020)]{2020arXiv200507210L}Liu, C., Fu, J., Shi, J., et al.\ 2020, arXiv:2005.07210
\bibitem[Maeder(2009)]{2009pfer.book.....M}Maeder, A.\ 2009, Physics, Formation and Evolution of Rotating Stars: , Astronomy and Astrophysics Library. ISBN 978-3-540-76948-4. Springer Berlin Heidelberg, 2009
\bibitem[Maeder et al.(2008)]{2008A&A...479L..37M}Maeder, A., Georgy, C., \& Meynet, G.\ 2008, \aap, 479, L37
\bibitem[Maeder et al.(1999)]{1999A&A...346..459M}Maeder, A., Grebel, E.~K., \& Mermilliod, J.-C.\ 1999, \aap, 346, 459
\bibitem[Maeder \& Meynet(2000)]{2000ARA&A..38..143M}Maeder, A. \& Meynet, G.\ 2000, \araa, 38, 143
\bibitem[Martayan et al.(2007a)]{2007A&A...472..577M}Martayan, C., Floquet, M., Hubert, A.~M., et al.\ 2007a, \aap, 472, 577
\bibitem[Martayan et al.(2006)]{2006A&A...452..273M}Martayan, C., Fr{\'e}mat, Y., Hubert, A.-M., et al.\ 2006, \aap, 452, 273
\bibitem[Martayan et al.(2007b)]{2007A&A...462..683M}Martayan, C., Fr{\'e}mat, Y., Hubert, A.-M., et al.\ 2007b, \aap, 462, 683
\bibitem[Meynet \& Maeder(2000)]{2000A&A...361..101M}Meynet, G. \& Maeder, A.\ 2000, \aap, 361, 101
\bibitem[Meynet et al.(2006a)]{2006isna.confE..15M}Meynet, G., Maeder, A., Hirschi, R., et al.\ 2006a, International Symposium on Nuclear Astrophysics - Nuclei in the Cosmos, 15.1
\bibitem[Meynet et al.(2006b)]{2006astro.ph.11261M}Meynet, G., Mowlavi, N., \& Maeder, A.\ 2006b, astro-ph/0611261
\bibitem[Milone et al.(2015)]{2015MNRAS.450.3750M}Milone, A.~P., Bedin, L.~R., Piotto, G., et al.\ 2015, \mnras, 450, 3750
\bibitem[Milone et al.(2016)]{2016MNRAS.458.4368M}Milone, A.~P., Marino, A.~F., D'Antona, F., et al.\ 2016, \mnras, 458, 4368
\bibitem[Milone et al.(2017)]{2017MNRAS.465.4363M}Milone, A.~P., Marino, A.~F., D'Antona, F., et al.\ 2017, \mnras, 465, 4363
\bibitem[Milone et al.(2018)]{2018MNRAS.477.2640M}Milone, A.~P., Marino, A.~F., Di Criscienzo, M., et al.\ 2018, \mnras, 477, 2640
\bibitem[Moe et al.(2019)]{2019ApJ...875...61M}Moe, M., Kratter, K.~M., \& Badenes, C.\ 2019, \apj, 875, 61
\bibitem[Nguyen et al.(2022)]{2022A&A...665A.126N}Nguyen, C.~T., Costa, G., Girardi, L., et al.\ 2022, \aap, 665, A126
\bibitem[Oksala et al.(2012)]{2012MNRAS.419..959O}Oksala, M.~E., Wade, G.~A., Townsend, R.~H.~D., et al.\ 2012, \mnras, 419, 959
\bibitem[Paunzen et al.(2021)]{2021A&A...645A..34P}Paunzen, E., H{\"u}mmerich, S., \& Bernhard, K.\ 2021, \aap, 645, A34
\bibitem[Pedersen(2022)]{2022ApJ...940...49P}Pedersen, M.~G.\ 2022, \apj, 940, 49
\bibitem[Perez \& Granger(2007)]{2007CSE.....9c..21P}Perez, F. \& Granger, B.~E.\ 2007, Computing in Science and Engineering, 9, 21
\bibitem[Pietrinferni et al.(2004)]{2004ApJ...612..168P}Pietrinferni, A., Cassisi, S., Salaris, M., et al.\ 2004, \apj, 612, 168
\bibitem[Pietrinferni et al.(2006)]{2006ApJ...642..797P}Pietrinferni, A., Cassisi, S., Salaris, M., et al.\ 2006, \apj, 642, 797
\bibitem[Potter et al.(2012)]{2012MNRAS.419..748P}Potter, A.~T., Tout, C.~A., \& Eldridge, J.~J.\ 2012, \mnras, 419, 748
\bibitem[Preston(1974)]{1974ARA&A..12..257P}Preston, G.~W.\ 1974, \araa, 12, 257
\bibitem[Privitera et al.(2016)]{2016A&A...591A..45P}Privitera, G., Meynet, G., Eggenberger, P., et al.\ 2016, \aap, 591, A45
\bibitem[Puls et al.(2008)]{2008A&ARv..16..209P}Puls, J., Vink, J.~S., \& Najarro, F.\ 2008, \aapr, 16, 209
\bibitem[Qin et al.(2019)]{2019ApJS..242...13Q}Qin, L., Luo, A.-L., Hou, W., et al.\ 2019, \apjs, 242, 13
\bibitem[Quentin \& Tout(2018)]{2018MNRAS.477.2298Q}Quentin, L.~G. \& Tout, C.~A.\ 2018, \mnras, 477, 2298
\bibitem[Raghavan et al.(2010)]{2010ApJS..190....1R}Raghavan, D., McAlister, H.~A., Henry, T.~J., et al.\ 2010, \apjs, 190, 1
\bibitem[Ramachandran et al.(2018)]{2018A&A...609A...7R}Ramachandran, V., Hainich, R., Hamann, W.-R., et al.\ 2018, \aap, 609, A7
\bibitem[Ramachandran et al.(2019)]{2019A&A...625A.104R}Ramachandran, V., Hamann, W.-R., Oskinova, L.~M., et al.\ 2019, \aap, 625, A104
\bibitem[Ram{\'\i}rez-Agudelo et al.(2013)]{2013A&A...560A..29R}Ram{\'\i}rez-Agudelo, O.~H., Sim{\'o}n-D{\'\i}az, S., Sana, H., et al.\ 2013, \aap, 560, A29
\bibitem[Reese et al.(2012)]{2012A&A...539A..63R}Reese, D.~R., Marques, J.~P., Goupil, M.~J., et al.\ 2012, \aap, 539, A63
\bibitem[Riello et al.(2021)]{2021A&A...649A...3R}Riello, M., De Angeli, F., Evans, D.~W., et al.\ 2021, \aap, 649, A3
\bibitem[Rosen et al.(2012)]{2012ApJ...748...97R}Rosen, A.~L., Krumholz, M.~R., \& Ramirez-Ruiz, E.\ 2012, \apj, 748, 97
\bibitem[Royer et al.(2007)]{2007A&A...463..671R}Royer, F., Zorec, J., \& G{\'o}mez, A.~E.\ 2007, \aap, 463, 671
\bibitem[Sackmann(1970)]{1970A&A.....8...76S}Sackmann, I.~J.\ 1970, \aap, 8, 76
\bibitem[Sana et al.(2012)]{2012Sci...337..444S}Sana, H., de Mink, S.~E., de Koter, A., et al.\ 2012, Science, 337, 444
\bibitem[Santos et al.(2021)]{2021ApJS..255...17S}Santos, A.~R.~G., Breton, S.~N., Mathur, S., et al.\ 2021, \apjs, 255, 17
\bibitem[Sanyal et al.(2017)]{2017A&A...597A..71S}Sanyal, D., Langer, N., Sz{\'e}csi, D., et al.\ 2017, \aap, 597, A71
\bibitem[Sim{\'o}n-D{\'\i}az \& Herrero(2014)]{2014A&A...562A.135S}Sim{\'o}n-D{\'\i}az, S. \& Herrero, A.\ 2014, \aap, 562, A135
\bibitem[Skrutskie et al.(2006)]{2006AJ....131.1163S}Skrutskie, M.~F., Cutri, R.~M., Stiening, R., et al.\ 2006, \aj, 131, 1163
\bibitem[Spruit(1999)]{1999A&A...349..189S}Spruit, H.~C.\ 1999, \aap, 349, 189
\bibitem[Stauffer \& Hartmann(1986)]{1986PASP...98.1233S}Stauffer, J.~B. \& Hartmann, L.~W.\ 1986, \pasp, 98, 1233
\bibitem[Sun et al.(2021a)]{2021ApJS..257...22S}Sun, W., Duan, X.-W., Deng, L., et al.\ 2021a, \apjs, 257, 22
\bibitem[Sun et al.(2021b)]{2021ApJ...921..145S}Sun, W., Duan, X.-W., Deng, L., et al.\ 2021b, \apj, 921, 145
\bibitem[Sun et al.(2019)]{2019ApJ...883..182S}Sun, W., Li, C., Deng, L., et al.\ 2019, \apj, 883, 182
\bibitem[Takahashi \& Langer(2021)]{2021A&A...646A..19T}Takahashi, K. \& Langer, N.\ 2021, \aap, 646, A19
\bibitem[Tian et al.(2023)]{2023ApJS..266...14T}Tian, X.-. man ., Wang, Z.-. hua ., Zhu, L.-. ying ., et al.\ 2023, \apjs, 266, 14
\bibitem[Ud-Doula et al.(2009)]{2009MNRAS.392.1022U}Ud-Doula, A., Owocki, S.~P., \& Townsend, R.~H.~D.\ 2009, \mnras, 392, 1022
\bibitem[van Belle(2012)]{2012A&ARv..20...51V}van Belle, G.~T.\ 2012, \aapr, 20, 51
\bibitem[Vink et al.(2001)]{2001A&A...369..574V}Vink, J.~S., de Koter, A., \& Lamers, H.~J.~G.~L.~M.\ 2001, \aap, 369, 574
\bibitem[Wang et al.(2023)]{2023A&A...674A.129W}Wang, C., Yuan, H., Xiang, M., et al.\ 2023, \aap, 674, A129
\bibitem[Weber \& Davis(1967)]{1967ApJ...148..217W}Weber, E.~J. \& Davis, L.\ 1967, \apj, 148, 217
\bibitem[Wolff et al.(2007)]{2007AJ....133.1092W}Wolff, S.~C., Strom, S.~E., Dror, D., et al.\ 2007, \aj, 133, 1092
\bibitem[Wolff et al.(2004)]{2004ApJ...601..979W}Wolff, S.~C., Strom, S.~E., \& Hillenbrand, L.~A.\ 2004, \apj, 601, 979
\bibitem[Xiang et al.(2022)]{2022A&A...662A..66X}Xiang, M., Rix, H.-W., Ting, Y.-S., et al.\ 2022, \aap, 662, A66
\bibitem[Yusof et al.(2022)]{2022MNRAS.511.2814Y}Yusof, N., Hirschi, R., Eggenberger, P., et al.\ 2022, \mnras, 511, 2814
\bibitem[Zahn(1992)]{1992A&A...265..115Z}Zahn, J.-P.\ 1992, \aap, 265, 115
\bibitem[Zhang et al.(2021)]{2021ApJS..256...14Z}Zhang, B., Li, J., Yang, F., et al.\ 2021, \apjs, 256, 14
\bibitem[Zhang et al.(2020)]{2020ApJS..246....9Z}Zhang, B., Liu, C., \& Deng, L.-C.\ 2020, \apjs, 246, 9
\bibitem[Zorec(2023)]{2023Galax..11...54Z}Zorec, J.\ 2023, Galaxies, 11, 54
\bibitem[Zorec \& Royer(2012)]{2012A&A...537A.120Z}Zorec, J. \& Royer, F.\ 2012, \aap, 537, A120
\bibitem[{\"O}pik(1951)]{1951MNRAS.111..278O}{\"O}pik, E.~J.\ 1951, \mnras, 111, 278
\end{thebibliography}
\end{document}